\documentclass[aip,cha,reprint]{revtex4-1}

\usepackage[pdftex]{graphicx}
\usepackage{amssymb,amsfonts,amsmath}

\begin{document}


\title{Digit replacement: A generic map for nonlinear dynamical systems} 



\author{Vladimir Garc\'{\i}a-Morales}
\email[]{garmovla@uv.es}
\affiliation{Departament de Termodin\`amica, Universitat de Val\`encia, E-46100 Burjassot, Spain}

\date{\today}

\begin{abstract}
A simple discontinuous map is proposed as a generic model for nonlinear dynamical systems. The orbit of the map admits exact solutions for wide regions in parameter space and the  method employed (digit manipulation) allows the mathematical design of useful signals, such as regular or aperiodic oscillations with specific waveforms, the construction of complex attractors with nontrivial properties as well as the coexistence of different basins of attraction in phase space with different qualitative properties. A detailed analysis of the dynamical behavior of the map suggests how the latter can be used in the modeling of complex nonlinear dynamics including, e.g., aperiodic nonchaotic attractors and the hierarchical deposition of grains of different sizes on a surface.
\end{abstract}

\pacs{02.70.Bf, 05.45.Ac, 05.45.Df}

\maketitle 

\begin{quotation}
The quest for simple deterministic models that exhibit complex and unpredictable temporal evolution\cite{May,PeitgenBOOK,Lichtenberg,Wolfram} has been the subject of intense interest. Most remarkable examples are provided by systems of nonlinear ordinary differential equations leading to chaotic behavior, as the Lorenz \cite{Lorenz} or the R\"ossler \cite{Rossler} systems, and by time-discrete difference equations, as the logistic map \cite{May, Feigenbaum} or the Bernoulli shift \cite{Lichtenberg}. All these systems involve dynamical variables taking on numerical values that can be expanded in terms of a (generally infinite) convergent series of rational numbers in a radix (base) $p \in \mathbb{N}$ ($p>1$). The latter are each formed by an integer power of the radix $p$ multiplied by an integer in the interval $[0,p-1]$ called a digit. In this work, by directly addressing and manipulating digits (by means of a digit function that we have recently introduced \cite{CHAOSOLFRAC,PHYSAFRAC,semipredo,EPL}) we uncover a digit replacement operator that allows a generic map for nonlinear dynamical systems to be formulated. Our model is exactly solvable in wide regions of parameter space and encompasses a wide variety of complex nonlinear dynamical behavior from a new perspective.
\end{quotation}


\section{Introduction}
%

A dynamical system \cite{Lind,RuelleEckmann} is a pair $(M, \phi)$, where $M$ is a compact metric space and $\phi$ a continuous map $\phi: M \to M$. In this article we introduce a generic map for nonlinear dynamical systems $\phi: \mathbb{R} \to \mathbb{R}$ where $\phi$ is discontinuous. However, as we explicitly show, discontinuity can be seen here as arising from the discretization (shadowing) of any arbitrary specific continuous dynamics as governed by an arbitrary non-autonomous map. Some advantages of our formulation become evident, as soon as it is realized that the mathematical method presented allows each digit of any numerical quantity to be individually addressed. As a consequence of this, the temporal evolution that constitutes the output of the map can be mathematically designed in detail \emph{a priori} through its few control parameters. We find, for example, that the specific shape of any regular oscillations on an attracting limit cycle can be mathematically engineered. The method also allows the design/modeling of time series with irregular or aperiodic waveforms.


This article is organized as follows. In Section \ref{digit} we introduce the mathematical tools, the digit function and the replacement operator, in which our approach is based. Then, we show how to convert any arbitrary non-autonomous map into the generic map that is the object of our study (see Eq. (\ref{absmap}) below). In Section \ref{maphor} we discuss in detail its dynamics,  providing explicit solutions for the orbits of the map. Our discussion culminates in a model for aperiodic nonchaotic attractors, showing how they can be constructed in phase space, all attractors sharing aspects of the same qualitative dynamics (even when they can be made distinguishable). We prove that, in spite of the dynamics on the attractors being aperiodic, the Lyapunov exponent is zero.

%

\section{Formulation of the map} \label{digit}

Since our approach is based on the possibilities opened by a digit function that we have recently discussed in connection with fractals \cite{CHAOSOLFRAC, PHYSAFRAC,VGM4} and cellular automata \cite{semipredo,EPL} we first introduce this function. Let $p>1$ be a natural number called the \emph{radix}. Any arbitrary real number $x$ can be represented in radix $p$ as \cite{CHAOSOLFRAC}
\begin{equation}
x=\text{sign}(x)\sum_{k=-\infty}^{\lfloor \log_{p}|x| \rfloor} p^{k} \mathbf{d}_{p}(k,|x|) \label{iden}
\end{equation}
where $\mathbf{d}_{p}(k,x)$ is the digit function, defined as \cite{CHAOSOLFRAC}
\begin{equation}
\mathbf{d}_{p}(k,x)=\left \lfloor \frac{x}{p^{k}} \right \rfloor-p\left \lfloor \frac{x}{p^{k+1}} \right \rfloor  \label{cucuA}
\end{equation}
and where the brackets $\left \lfloor \ldots \right \rfloor$ denote the floor (lower closest integer) function. As an example, in the decimal radix $p=10$, the number $\pi \equiv 3.1415\ldots$ has digits $\mathbf{d}_{10}(0,\pi)=3$, $\mathbf{d}_{10}(-1,\pi)=1$, $\mathbf{d}_{10}(-2,\pi)=4$, $\mathbf{d}_{10}(-3,\pi)=1$, $\mathbf{d}_{10}(-4,\pi)=5$ etc.

Trivial properties of the digit function that can be directly checked from its definition are the $p^{k+1}$ periodicity
\begin{equation}
\mathbf{d}_{p}(k,x+p^{k+1})=\mathbf{d}_{p}(k,x) \label{periodigit}
\end{equation} 
and the scaling property
\begin{equation}
\mathbf{d}_{p}(k,x)=\mathbf{d}_{p}(0,p^{-k}x) \label{scaldigit}
\end{equation} 
The reader is referred to \cite{CHAOSOLFRAC} for other properties of the digit function and a detailed discussion/derivation of them. 

In this paper we shall consider $t$ a non-negative integer. In this case $\mathbf{d}_{p}(0,t)=t \mod p$, i.e. the zeroth digit function of an integer number is equal to that number modulo the radix $p$. We then observe that 
\begin{eqnarray}
\mathbf{d}_{1}(0,t)&=&0 \label{case1} \\
\mathbf{d}_{\infty}(0,t)&=&t \label{case2}
\end{eqnarray}

The number $\lfloor \log_{p}|x| \rfloor$ in Eq. (\ref{iden}) is the maximum exponent of the power of the radix with which $x \in \mathbb{R}$ is represented in radix $p$. \cite{VGM4} Thus, the function $\lfloor \log_{p}|x| \rfloor$ gives the \emph{order of magnitude} of $x$ when $x$ is written in radix $p$. The following function
\begin{equation}
\text{lp}_{p}(x)\equiv\text{sign}(x)p^{\lfloor \log_{p}|x| \rfloor}\mathbf{d}_{p}(\lfloor \log_{p}x \rfloor,x) \label{leadp}
\end{equation}
gives the \emph{leading part} of $x$, when written in radix $p$. For example, let $x= 6.022\cdot10^{23}$, then $\text{lp}_{10}(x)=6\cdot 10^{23}$. As another example, we have $\text{lp}_{10}(204)=2\cdot 10^{2}$ but $\text{lp}_{2}(204)=1\cdot 2^{7}$
since 204 is written in radix $p=2$ as $(11001100)_{2}$ and thus $204=1\cdot 2^{7}+1\cdot 2^{6}+0\cdot 2^{5}+0\cdot 2^{4}+1\cdot 2^{3}+1\cdot 2^{2}+0\cdot 2^{1}+0\cdot 2^{0}$.
 
The map studied in this article has the general form
\begin{equation}
x_{t+1}=x_{t}+p^{j_{t}}\left(\mathbf{d}_{p}(k_{t},\mu_{t})-\mathbf{d}_{p}(j_{t},x_{t})\right)  \label{absmap}
\end{equation}
where $x_{t} \in \mathbb{R}$ and $\mu_{t} \in \mathbb{R}$ are both non-negative real-valued functions at discrete time $t$, $x_{t}$ being the dynamical variable and $\mu_{t}$ being the `objective signal' ruling the instantaneous dynamics and $j_{t}$ and $k_{t}$ are integer-valued functions that also evolve with time. The parameter $p$, called the radix, is a natural number $p\ge 2$. 


The map Eq. (\ref{absmap}) has the following meaning: At time $t+1$, the value of the dynamical variable $x_{t+1}$ is given by $x_{t}$ but replacing the $j_{t}$-th digit of $x_{t}$ by the $k_{t}$-th digit of $\mu_{t}$ when both numbers are written in radix $p$. Thus, the map replaces a single digit of the numerical value of $x_{t}$ at each time step by a single digit of $\mu_{t}$.


To show how Eq. (\ref{absmap}) can be derived starting from an arbitrary non-autonomous map
\begin{equation}
x_{t+1}=F(x_{t}, t) \label{nonauto}
\end{equation}
we approximate $x_{t}$ at each instant of time $t$ by the leading part of $x_{t}$ (i.e. we neglect the contribution to $x_{t}$ of any digit that is not the most significant one). Therefore, we have, in terms of the leading part $\text{lp}_{p}(x)$ function defined in Eq. (\ref{leadp}), 
\begin{eqnarray}
x_{t} &\approx& \text{lp}_{r_{t}}(x_{t}) \label{eq1} \\
x_{t+1} &\approx& \text{lp}_{q_{t}}(x_{t+1})=\text{lp}_{q_{t}}\left(F(x_{t}, t)\right) \label{eq2}
\end{eqnarray}
for any $r_{t}, q_{t} \ge 2$. The radix values $r_{t}, q_{t} \in \mathbb{N}$ can be conveniently chosen at each time so as to optimize the approximation \cite{QUANTUM}. We demand here that they are equal to some powers $r_{t}\equiv p^{h_{t}^{(1)}}$, $q_{t}\equiv p^{h_{t}^{(2)}}$ of a natural number $p\ge 2$, $h_{t}^{(1)}$ and $h_{t}^{(2)}$ being integers. We let $p$ constant and regard $h_{t}^{(1)}$ and $h_{t}^{(2)}$ as functions of time. We also demand the following inequalities (which can always be satisfied by suitably chosing $p$) 
\begin{eqnarray}
&&h_{t}^{(1)}\left \{ \frac{\log_{p}x_{t}}{h_{t}^{(1)}} \right \} < 1 \nonumber \\
&&h_{t}^{(2)}\left \{ \frac{\log_{p}F(x_{t}, t)}{h_{t}^{(2)}} \right \} <1 \nonumber
\end{eqnarray}
where the brackets $\{\ldots \}$ denote the fractional part function. By subtracting Eqs. (\ref{eq1}) and (\ref{eq2}), we obtain
\begin{eqnarray}
&&x_{t+1}\approx x_{t}+\text{lp}_{q_{t}}\left(F(x_{t}, t)\right)-\text{lp}_{r_{t}}(x_{t}) \nonumber \\
&&= x_{t}+q_{t}^{\lfloor \log_{q_{t}}F(x_{t}, t) \rfloor}\mathbf{d}_{q_{t}}(\lfloor \log_{q_{t}}F(x_{t}, t) \rfloor,F(x_{t}, t))-\nonumber \\
&&\ \ \ -r_{t}^{\lfloor \log_{r_{t}}x_{t} \rfloor}\mathbf{d}_{r_{t}}(\lfloor \log_{r_{t}}x_{t} \rfloor,x_{t}) \nonumber \\
&&= x_{t}+p^{h_{t}^{(2)}\lfloor \log_{q_{t}}F(x_{t}, t) \rfloor}\mathbf{d}_{q_{t}}(\lfloor \log_{q_{t}}F(x_{t}, t) \rfloor,F(x_{t}, t))-\nonumber \\
&&\ \ \ -p^{h_{t}^{(1)}\lfloor \log_{r_{t}}x_{t} \rfloor}\mathbf{d}_{r_{t}}(\lfloor \log_{r_{t}}x_{t} \rfloor,x_{t})  \label{comeon} 
\end{eqnarray}
We now note that, 
\begin{eqnarray}
\mathbf{d}_{r_{t}}(\lfloor \log_{r_{t}}x_{t} \rfloor,x_{t})&=&\mathbf{d}_{p^{h_{t}^{(1)}}}(\lfloor \log_{r_{t}}x_{t} \rfloor,x_{t}) \nonumber \\
&=&\mathbf{d}_{p}(h_{t}^{(1)}\lfloor \log_{r_{t}}x_{t} \rfloor,x_{t}) \label{comeon2}
\end{eqnarray}
as well as $\mathbf{d}_{q_{t}}(\lfloor \log_{q_{t}}F(x_{t}, t) \rfloor,F(x_{t}, t))=\mathbf{d}_{p}(h_{t}^{(2)}\lfloor \log_{q_{t}}F(x_{t}, t) \rfloor,F(x_{t}, t))$. Thus, if we take $h_{t}^{(1)}$ and $h_{t}^{(2)}$ at each time so that
\begin{equation}
h_{t}^{(2)}\left \lfloor \frac{\log_{p}F(x_{t}, t)}{h_{t}^{(2)}} \right \rfloor-h_{t}^{(1)}\left \lfloor \frac{\log_{p}x_{t}}{h_{t}^{(1)}} \right \rfloor=0 \label{demand}
\end{equation}
we see that, if we identify
\begin{eqnarray}
k_{t}&\equiv& h_{t}^{(2)}\left \lfloor \frac{\log_{p}F(x_{t}, t)}{h_{t}^{(2)}} \right \rfloor \\
j_{t}&\equiv& h_{t}^{(1)}\left \lfloor \frac{\log_{p}x_{t}}{h_{t}^{(1)}} \right \rfloor \\
\mu_{t} &\equiv& F(x_{t}, t)
\end{eqnarray}
by using Eqs. (\ref{comeon2}) and (\ref{demand}) and by replacing all these quantities in Eq. (\ref{comeon}) we finally obtain Eq. (\ref{absmap}). We note that there is considerable freedom above to choose $h_{t}^{(1)}$, $h_{t}^{(2)}$ and $\mu_{t}$. This freedom is related to a scaling property of the replacement dynamics that we point out below, see Eq. (\ref{rp0}). 


We thus conclude that \emph{the dynamics of any arbitrary non-autonomous map, Eq. (\ref{nonauto}), can always be approximated by a digit replacement dynamics given by Eq. (\ref{absmap})}. The contrary, however, is not necessarily true, i.e. \emph{there are choices for $j_{t}$, $k_{t}$ and $\mu_{t}$ in Eq. (\ref{absmap}) such that the latter equation cannot be approximated by Eq. (\ref{nonauto})}.

Since the digit replacement dynamics, Eq. (\ref{absmap}), is more general than the dynamics provided by a non-autonomous map of the form Eq. (\ref{nonauto}), the aim of the rest of this article is to study the digit replacement dynamics for the simplest input functions $j_{t}$ and $k_{t}$; the latter being constant, linear, or simply periodic functions of $t$. \emph{We shall take $\mu_{t}=\mu$ constant throughout.}

We define the replacement operator, $\mathcal{R}_{p}(j,k;a,b)$ as
\begin{equation}
\mathcal{R}_{p}(j,k;a,b)\equiv a+\text{sign}(a)p^{j}\left(\mathbf{d}_{p}(k,|b|)-\mathbf{d}_{p}(j,|a|)\right) \label{replao}
\end{equation}
This operator replaces the $j$-th digit of $a$ by the $k$-th digit of $b$ (when both are written in radix $p$). Thus, the digit replacement dynamics of map Eq. (\ref{absmap}) is concisely written in terms of this operator as
\begin{equation}
x_{t+1}=\mathcal{R}_{p}(j_{t},k_{t};x_{t},\mu_{t}) \label{thedyna}
\end{equation}

We observe that, for any integers $h$ and $j$, because of the scaling property of the digit function, Eq. (\ref{scaldigit}), the following important relationship holds
\begin{equation}
\mathcal{R}_{p}\left(h+j,i+k; p^{h}a, p^{i}b\right)=p^{h}\mathcal{R}_{p}\left(j,k; a, b \right) \label{rp0}
\end{equation}

\section{Analysis of the dynamics}\label{maphor}

The richness of the dynamical behavior of Eq. (\ref{thedyna}) is better explored and understood step by step starting from the most trivial instances of $j_{t}$ and $k_{t}$ and progressing to more complex behavior. If $j_{t}=\left \lfloor \log_{p} x_{t} \right \rfloor$ and $k_{t}=\left \lfloor \log_{p} \mu \right \rfloor$, with $\mu$ constant, Eq. (\ref{thedyna}) reduces to
\begin{equation}
x_{t+1}= \mathcal{R}_{p}\left(\left \lfloor \log_{p} x_{t} \right \rfloor,\left \lfloor \log_{p} \mu \right \rfloor; x_{t}, \mu \right) \label{thedyna0}
\end{equation}
which trivially replaces the most significant digit of $x_{t}$ by the most significant one of $\mu$ and does nothing else. Thus, already on the first time step, a fixed point $x_{\infty}$ is so reached, with value
\begin{equation}
x_{\infty}=x_{1}=\mathcal{R}_{p}\left(\left \lfloor \log_{p} x_{0} \right \rfloor,\left \lfloor \log_{p} \mu \right \rfloor; x_{0}, \mu \right) \label{thedyna0fp}
\end{equation}
In the following, we shall, of course, consider more interesting and increasingly complex dynamics.

\subsection{Relaxation to a fixed point}

\begin{figure*} 
\includegraphics[width=1.0\textwidth]{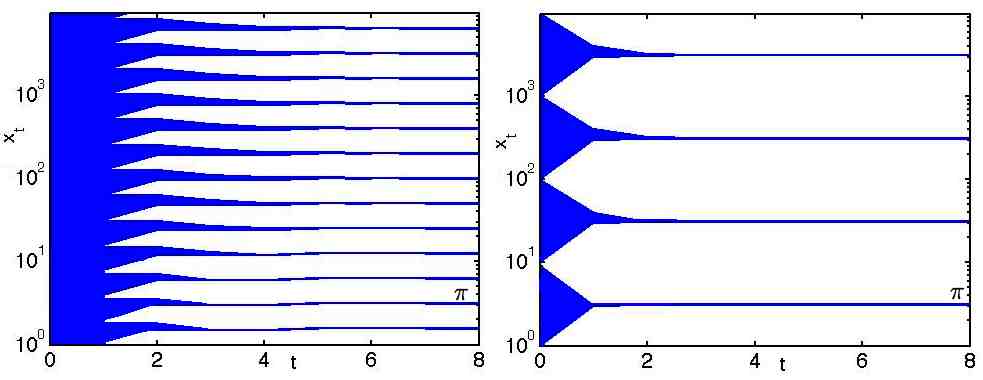}
\caption{\scriptsize{First eight iterates of the map Eq. (\ref{map1}) for $\mu=\pi$ starting from  initial conditions $x_{0} \in [1,9900]$ and for $p=2$ (left) and $p=10$ (right). The fixed point at $\mu$ ($=\pi$ in this case) is always present, independently of the value of $p$, although its basin of attraction is wider for $p$ larger. Note the logarithmic scale on the $x_{t}$ axis.}} \label{fpm}
\end{figure*}

We can now consider $j_{t}$ and $k_{t}$ as a digit replacement dynamics going from the most significant digits of $x_{t}$ and $\mu$ to the less ones
\begin{eqnarray}
j_{t} &\equiv& \left \lfloor \log_{p} x_{t} \right \rfloor-t \\
k_{t} &\equiv& \left \lfloor \log_{p} \mu \right \rfloor-t
\end{eqnarray}
so that Eq. (\ref{thedyna}) reduces to the simple form
\begin{equation}
x_{t+1}=\mathcal{R}_{p}\left(\left \lfloor \log_{p} x_{t} \right \rfloor-t,\left \lfloor \log_{p} \mu \right \rfloor-t; x_{t}, \mu\right) \label{map1}
\end{equation}
This simple map can exactly be solved for the orbit for any initial condition $x_{0}$ as
\begin{eqnarray}
x_{t}&=&\sum_{j=-\left \lfloor \log_{p} x_{0} \right \rfloor}^{t-1-\left \lfloor \log_{p} x_{0} \right \rfloor}p^{-j}\mathbf{d}_{p}(\left \lfloor \log_{p} \mu \right \rfloor-\left \lfloor \log_{p} x_{0} \right \rfloor-j,\mu)+ \nonumber \\
&&\ +\sum_{j=t-\left \lfloor \log_{p} x_{0} \right \rfloor}^{\infty}p^{-j}\mathbf{d}_{p}(-j,x_{0}) \label{orbit}
\end{eqnarray}
The proof of this result is trivial by induction. At $t=0$ we have
\begin{equation}
x_{t=0}=\sum_{j=-\left \lfloor \log_{p} x_{0} \right \rfloor}^{\infty}p^{-j}\mathbf{d}_{p}(-j,x_{0})=x_{0}
\end{equation}
Let us assume Eq. (\ref{orbit}) valid at time $t$. Then, at time $t+1$
\begin{eqnarray}
x_{t+1}&=& \mathcal{R}_{p}\left(\left \lfloor \log_{p} x_{t} \right \rfloor-t,\left \lfloor \log_{p} \mu \right \rfloor-t; x_{t}, \mu\right) \nonumber 
\end{eqnarray}
\begin{eqnarray}
&=&\sum_{j=-\left \lfloor \log_{p} x_{0} \right \rfloor}^{t-1-\left \lfloor \log_{p} x_{0} \right \rfloor}p^{-j}\mathbf{d}_{p}(\left \lfloor \log_{p} \mu \right \rfloor-\left \lfloor \log_{p} x_{0} \right \rfloor-j,\mu)+\nonumber \\
&&+\sum_{j=t-\left \lfloor \log_{p} x_{0} \right \rfloor}^{\infty}p^{-j}\mathbf{d}_{p}(-j,x_{0})+\nonumber \\
&&+p^{\left \lfloor \log_{p} x_{t} \right \rfloor-t}\left(\mathbf{d}_{p}(\left \lfloor \log_{p} \mu \right \rfloor-t,\mu)-\mathbf{d}_{p}(\left \lfloor \log_{p} x_{t} \right \rfloor-t,x_{0})\right) \nonumber \\
&=&\sum_{j=-\left \lfloor \log_{p} x_{0} \right \rfloor}^{t-\left \lfloor \log_{p} x_{0} \right \rfloor}p^{-j}\mathbf{d}_{p}(\left \lfloor \log_{p} \mu \right \rfloor-\left \lfloor \log_{p} x_{0} \right \rfloor-j,\mu)+ \nonumber \\
&&+\sum_{j=t+1-\left \lfloor \log_{p} x_{0} \right \rfloor}^{\infty}p^{-j}\mathbf{d}_{p}(-j,x_{0})
\label{orbitproof}
\end{eqnarray}
where we have used the obvious fact that, in this case $\left \lfloor \log_{p} x_{t} \right \rfloor=\left \lfloor \log_{p} x_{0} \right \rfloor$ $\forall t$. This proves the validity of Eq. (\ref{orbit}). Thus, the map converges to the fixed point 
\begin{eqnarray}
x_{\infty}&=&\sum_{j=-\left \lfloor \log_{p} x_{0} \right \rfloor}^{\infty}p^{-j}\mathbf{d}_{p}(\left \lfloor \log_{p} \mu \right \rfloor-\left \lfloor \log_{p} x_{0} \right \rfloor-j,\mu) \nonumber \\
&=&\mu p^{\left \lfloor \log_{p} x_{0} \right \rfloor-\left \lfloor \log_{p} \mu \right \rfloor}  \label{fpoints}
\end{eqnarray}
We observe that there is indeed an infinite number of isolated fixed points separated in value by powers of $p$. The fixed point $x_{\infty}=\mu$ is always present regardless of the value of $p$, although its basin of attraction (which corresponds to all those initial conditions for which $\left \lfloor \log_{p} x_{0} \right \rfloor=\left \lfloor \log_{p} \mu \right \rfloor$) is explicitly $p$-dependent. Thus, if $x_{\infty}$ is the fixed point corresponding to $x_{0}$, $x_{\infty}$ attracts all initial conditions in the semiopen interval $[p^{\left \lfloor \log_{p} x_{0} \right \rfloor}, p^{\left \lfloor \log_{p} x_{0} \right \rfloor+1})$ but there is also a fixed point $p^{n}x_{\infty}$ for any $n\in \mathbb{Z}$ which attracts all initial conditions in the interval $[p^{n+\left \lfloor \log_{p} x_{0} \right \rfloor}, p^{n+\left \lfloor \log_{p} x_{0} \right \rfloor+1})$. \emph{The existence of these fixed points is a consequence of the scale invariance implied by Eq. (\ref{rp0}).} From Eq. (\ref{fpoints}) we also observe that \emph{the larger $p$, the wider the basins of attractions of the fixed points}. We also note that for $x_{0}=0$, $x_{\infty}=0$ independently of $\mu$ since $\left \lfloor \log_{p} 0 \right \rfloor=-\infty$.

In Fig. \ref{fpm} the evolution of $x_{t}$ is plotted for initial conditions $x_{0} \in [1,9900]$ as obtained from Eq. (\ref{orbit}) for $\mu=\pi$ and for $p=2$ (left) and $p=10$ (right). We observe the existence of fixed points for the values predicted by Eq. (\ref{fpoints}) and the observation made above that the larger $p$, the wider their basins of attraction. In the range of initial conditions considered, there are 13 fixed points for $p=2$ and four fixed points for $p=10$. As we see, independently of the value of $p$, $\mu=\pi$ is always a fixed point for the range of initial conditions $\left \lfloor \log_{p} x_{0} \right \rfloor=\left \lfloor \log_{p} \mu \right \rfloor$ (the thickness of the basin of attraction of the fixed point being $p$-dependent) as obtained from the theory above. We also observe that the smaller $p$, the slower the convergence to the fixed points. This is clear from the fact that for lower $p$ and constant $x_{0}$, more digits are needed to render $x_{0}$ up to a certain precision and the number of such significant digits is directly proportional to the duration of the transient. The evolution to $\mu$, starting from an arbitrary $x_{0}$ generally converges in a non-monotonic, nonlinear manner. For $x_{0}=\sqrt{2}$ the successive iterates of the map for $\mu=\pi$ and $p=10$ are
\begin{eqnarray}
x_{0}=1.414213\ldots \ &\qquad& x_{1}=\mathbf{3}.414213\ldots \nonumber \\
x_{2}=\mathbf{3.1}14213\ldots \   & & x_{3}=\mathbf{3.14}4213\ldots \nonumber \\
x_{4}=\mathbf{3.141}213\ldots  &\qquad& x_{5}=\mathbf{3.1415}13\ldots \qquad  \nonumber \\
 \ldots \qquad x_{\infty}&=& \pi  \nonumber 
\end{eqnarray}
where it is clear that the convergence to the fixed point $\pi$ is achieved by replacing at each time step $t$ a less significant digit of $x_{0}$ by the corresponding one of $\pi$ (the digits that have been replaced are written in bold above). For $p=2$, we have, however, the sequence of iterates given by
\begin{eqnarray}
x_{0}=1.414213\ldots \ &\qquad& x_{1}=1.914213\ldots \nonumber \\
x_{2}=1.664213\ldots \   & & x_{3}=1.539213\ldots \nonumber \\
x_{4}=1.601713\ldots \ &\qquad& x_{5}=1.570463\ldots \qquad  \nonumber \\
 \ldots \qquad x_{\infty}&=&\pi/2  \nonumber 
\end{eqnarray}
as predicted by the above development. Although it is not apparent that this latter evolution is a digit replacement dynamics, this is only so because we are giving the numbers in the decimal radix. Had we chosen the binary radix for the representation of the above numbers, we would then see that, indeed, digits of $x_{t}$ (that are all 0 or 1) are being gradually replaced by those of $\mu$ (from the most to the less significant places). Thus, although the choice of the radix is immaterial for the numbers themselves (a number is `always the same' regardless of the radix used) the radix $p$, in which the dynamics governed by Eq. (\ref{map1}) takes place, constitutes the optimal \emph{representation} \cite{QUANTUM} to understand what is going on.

\begin{figure*} 
\includegraphics[width=1.0\textwidth]{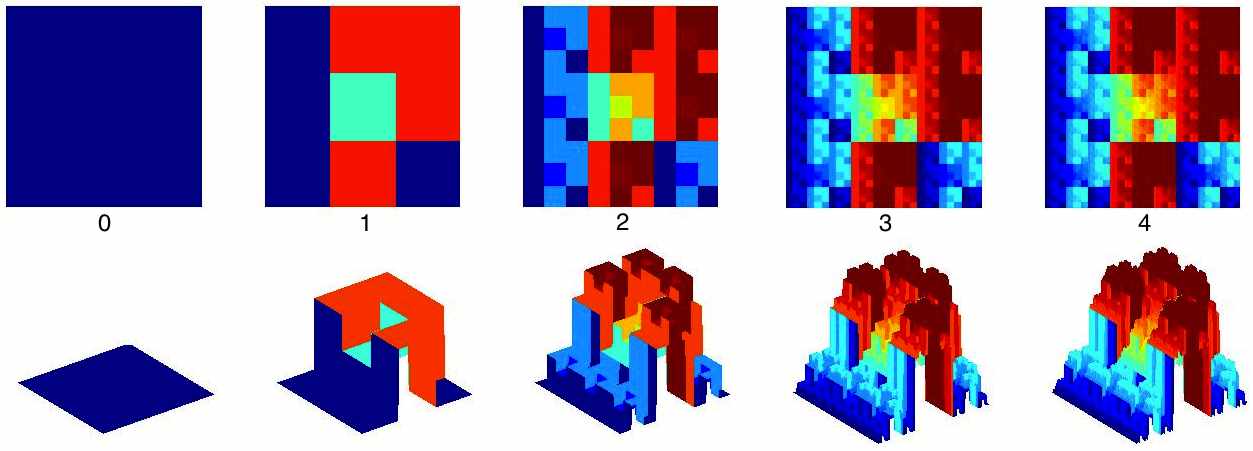}
\caption{\scriptsize{Four time steps of the spatiotemporal evolution $x_{t}(u,v)$ obtained from Eq. (\ref{mapbw}) for $p=3$ $R=6453$ $u \in [0,240]$ and $v \in [0,240]$. Two different views of the developing surface are shown, starting from a clean surface at $t=0$.}} \label{growf}
\end{figure*} 

Eq. (\ref{map1}) and the above mathematical development finds a physical application in the problem of hierarchical deposition of debris on a surface, in which grains of different sizes are deposited in an order related to their size, from larger to smaller grains \cite{Indekeu5, Indekeu4, Indekeu6, StanleyFractals}. Here, the deposition takes place according to a certain rule  in which grains of bigger size deposit first  and a next generation of grains, which are smaller by a factor $\sim 1/p^{3}$, deposit next. We first construct a model for $\mu$, that characterizes the objective attractor. We have recently found \cite{PHYSAFRAC}, that if $u$ and $v$ denote spatial coordinates on the plane $\mathbb{R}^{2}$ then, for $R \in [0,p^{p^{2}}-1]$ an integer and $k_{max}\equiv \max \{\lfloor \log_{p}u \rfloor, \lfloor \log_{p}v \rfloor \}$, the function given by
\begin{equation}
\mathbf{b}_{p}(u,v; R)=\sum_{k=-\infty}^{k_{max}} p^{k} \mathbf{d}_{p}\left(\mathbf{d}_{p}(k,u)+p\mathbf{d}_{p}(k,v), R \right) \label{bwq}
\end{equation}
is a surface with fractal self-affine properties \cite{PHYSAFRAC}. Thus, if we start from a `clean' surface $x_{0}(u,v)=0$ it is now clear, from the discussion given above, that this surface will be an attractor of the dynamics given by the map
\begin{eqnarray}
x_{t+1}(u,v)&=&\mathcal{R}_{p}\left(\left \lfloor \log_{p} \mathbf{b}_{p}(u,v; R) \right \rfloor-t,\right. \nonumber \\
&&\qquad \left. \left \lfloor \log_{p} \mathbf{b}_{p}(u,v; R) \right \rfloor-t;\right. \nonumber \\
&&\qquad \left. x_{t}(u,v), \mathbf{b}_{p}(u,v; R) \right) \label{mapbw}
\end{eqnarray}
which corresponds to Eq. (\ref{map1}) with $\mu=\mathbf{b}_{p}(u,v; R)$ and with $x_{t}$ having now a spatial dependence $x_{t}(u,v)$ on the real coordinates $u$ and $v$. We also fix, in Eq. (\ref{map1}), $\left \lfloor \log_{p} x_{t} \right \rfloor=\left \lfloor \log_{p} x_{0} \right \rfloor=\left \lfloor \log_{p} \mathbf{b}_{p}(u,v; R) \right \rfloor$. Thus, from Eq. (\ref{fpoints}), we observe that if we start from a `clean' surface $x_{0}(u,v)=0$ the map shall converge to 
\begin{equation}
x_{\infty}(u,v)=\mathbf{b}_{p}(u,v; R)
\end{equation}
for any point $(u,v)$ on the plane. This convergence proceeds as follows. On the clean surface, in the first iteration the first grains are deposited, with area $p^{\left \lfloor \log_{p} \mathbf{b}_{p}(u,v; R) \right \rfloor} \times p^{\left \lfloor \log_{p} \mathbf{b}_{p}(u,v; R) \right \rfloor}$ and height $p^{\left \lfloor \log_{p} \mathbf{b}_{p}(u,v; R) \right \rfloor}\mathbf{d}_{p}\left(\mathbf{d}_{p}(k_{max},u)+p\mathbf{d}_{p}(k_{max},v), R \right)$. Then, in the next time step smaller grains with area $p^{\left \lfloor \log_{p} \mathbf{b}_{p}(u,v; R) \right \rfloor-1} \times p^{\left \lfloor \log_{p} \mathbf{b}_{p}(u,v; R) \right \rfloor-1}$  are deposited. The process goes on from the bigger to the smaller scales until convergence to the fractal object $\mathbf{b}_{p}(u,v; R)$ is achieved. More significant digits of $\mathbf{b}_{p}(u,v; R)$ have a longer spatial variability as we prove in the coarse graining theorem recently reported in \cite{PHYSAFRAC}.

In Figure \ref{growf} this process of fractal surface growth is shown, as obtained from Eq. (\ref{mapbw}) for for $p=3$ $R=6453$ $u \in [0,240]$ and $v \in [0,240]$. Four time steps of the spatiotemporal evolution $x_{t}(u,v)$ are shown (the time step is indicated on each panel), giving two different views of the same surface. We observe a fractal surface deterministically growing from the coarser to the finer details. Although we have considered a digit replacement dynamics from most to less significant digits, it is also possible to select at random the digit to be replaced and the attractor will be the same, after a sufficiently large number of time steps. In combination with our results in \cite{PHYSAFRAC} this provides an explanation of why iterated functions systems \cite{Barnsley} generally lead to fractal attractors.




\subsection{Periodic orbits}

\begin{figure*} 
\includegraphics[width=1.0\textwidth]{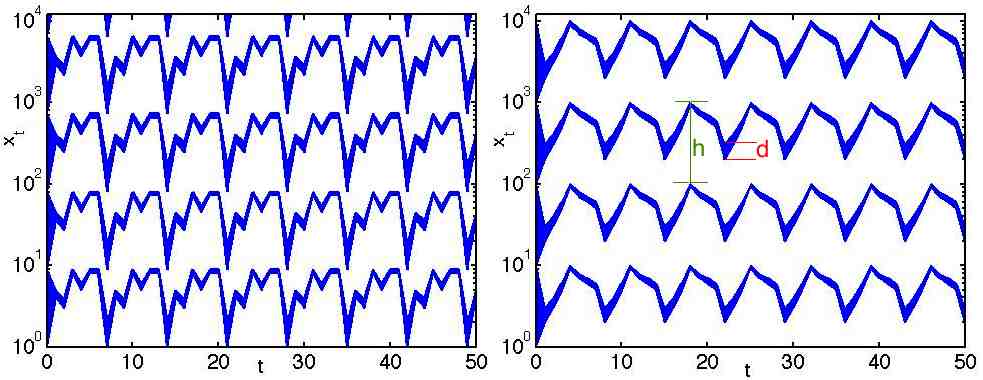}
\caption{\scriptsize{Evolution of $x_{t}$ obtained from the map Eq. (\ref{bmap}) for $p=9$ (left) and $p=10$ (right), for $\mu=2359765$ and $n=7$, starting from  initial conditions $x_{0} \in [1,9900]$. Note the logarithmic scale on the $x_{t}$ axis.}} \label{oscil1}
\end{figure*} 

Let us now consider $j_{t}$ and $k_{t}$ given by the following simple expressions
\begin{eqnarray}
j_{t} &\equiv& \left \lfloor \log_{p} x_{t} \right \rfloor \\
k_{t} &\equiv& \left \lfloor \log_{p} \mu \right \rfloor-\mathbf{d}_{n}(0,t)
\end{eqnarray}
in which case, Eq. (\ref{thedyna}), reduces to
\begin{equation}
x_{t+1}= \mathcal{R}_{p}\left(\left \lfloor \log_{p} x_{t} \right \rfloor,\left \lfloor \log_{p} \mu \right \rfloor-\mathbf{d}_{n}(0,t); x_{t}, \mu \right) \label{bmap}
\end{equation} 
At each time step the most significant digit of $x_{t}$ is replaced by a digit of $\mu$ that is $\mathbf{d}_{n}(0,t)$ digits away (to less significant places) from the most significant digit of $\mu$, given by  $\left \lfloor \log_{p} \mu \right \rfloor$. We note that, at every time, only one of the $n$ most significant digits of $\mu$ are used to replace the most significant digit of $x_{t}$. This is so because of the digit function being periodic, as made explicit by Eq. (\ref{periodigit}), and one has
\begin{equation}
\mathbf{d}_{n}(0,t+n)=\mathbf{d}_{n}(0,t) \label{Nperiod}
\end{equation} 
There are thus two possibilities: 1) any of the $n$ most significant digits of $\mu$ is equal to zero when written in radix $p$, in which case $x_{t}$ converges to zero; 2) none of the $n$ most significant digits of $\mu$ when written in radix $p$ is zero, in which case, because of Eq. (\ref{Nperiod}) the orbit of $x_{t}$ has period $n$ as well. 

The case 1) is easy to understand. For, if any of the $n$ most significant digits of $\mu$ in radix $p$ is zero, after $n$ time steps the most significant digit of $x_{t}$ has been once replaced by zero and as a result $\left \lfloor \log_{p} x_{t} \right \rfloor$ becomes more negative. This means that the leading part of $x_{t}$ decreases in a factor of $p$.
If this process is repeated an infinite number of times
$\left \lfloor \log_{p} x_{t} \right \rfloor=-\infty$ which means that the leading partis proportional to $p^{-\infty}$, i.e. that $x_{t}$ has converged to zero.

We now concentrate on the more interesting case 2) in which none of the $n$ most significant digits of $\mu$ is zero. The orbit can again be trivially solved in this simple case, and has for any $t \ge 1$ the form
\begin{eqnarray}
x_{t}&=&\mathcal{R}_{p}\left(\left \lfloor \log_{p} x_{0} \right \rfloor,\left \lfloor \log_{p} \mu \right \rfloor-\mathbf{d}_{n}(0,t-1); x_{0}, \mu \right)+ \nonumber \\
&&+\sum_{j=0}^{\left \lfloor \log_{p} x_{0} \right \rfloor-1}p^{-j}\mathbf{d}_{p}(-j,x_{0}) \label{orbitos}
\end{eqnarray}
since all time dependence affects only the most significant digit of $x_{t}$ accompanying a power $p^{\left \lfloor \log_{p} x_{0} \right \rfloor}$ of the radix and all other digits are left invariant. This time dependence has period $n$ for all $t>0$ as can be seen by making the transformation $t \to t+n$ and observing that, since $\mathbf{d}_{n}(0,t-1+n)=\mathbf{d}_{n}(0,t-1)$, this implies, from Eq. (\ref{orbitos}) that $x_{t}=x_{t+n} \forall t \ge 1$. Thus, for example, for $b=27$, $p=10$ and starting from $x_{0}=\sqrt{2}$, Eq. (\ref{bmap}) produces, for $p=10$ the iterates
\begin{eqnarray}
x_{0}&=&1.41\ldots \qquad
x_{1}=2.41\ldots \qquad
x_{2}=7.41\ldots \qquad \nonumber \\
x_{3}&=&2.41\ldots \qquad x_{4}=7.41\ldots \qquad \ldots \nonumber 
\end{eqnarray}


\begin{figure*} 
\includegraphics[width=1.0\textwidth]{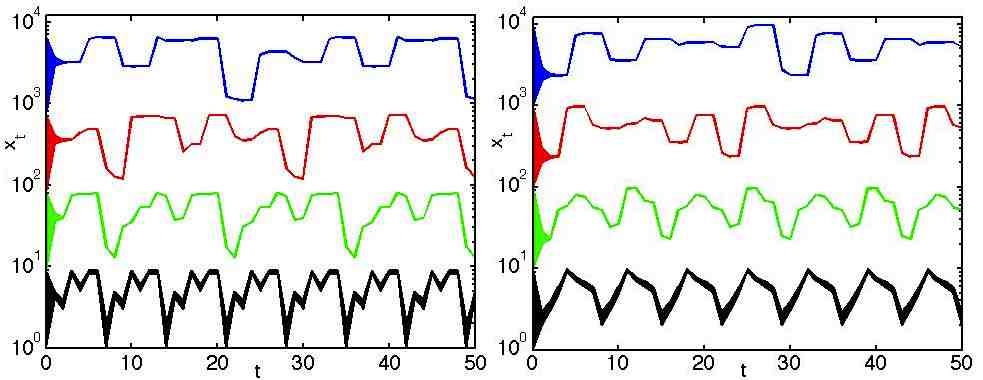}
\caption{\scriptsize{Temporal evolution of $x_{t}$ obtained from the map Eq. (\ref{thedyna}) for $p=9$ (left) and $p=10$ (right), for $\mu=2359765$, $n=7$ and $m=1+|\left \lfloor \log_{p} x_{0} \right \rfloor|$. Note the logarithmic scale on the $x_{t}$ axis.}} \label{oscil2}
\end{figure*}

In Fig. \ref{oscil1}, the evolution of the map Eq. (\ref{bmap}) is plotted for $p=9$ (left) and $p=10$ (right), for $\mu=2359765$ and $n=7$. In both cases the orbits have period $T=n=7$ for all initial conditions. Although the period is the same, the shape of the oscillations is affected by changing $p$. Note that for $p=10$ the digits of $\mu=2359765$ from the most significant to the less significant obey: $2 < 3 < 5 < 9$, $9> 7 > 6 > 5$. As a consequence, the shape of the oscillation consist in each period of two different monotonic parts, an increasing one and a decreasing one. For $p=9$, however, since $\mu=2359765$ is written as $\mu=(4385881)_{9}$ in radix $9$, we have that the shape of the oscillation consists of $5$ different monotonic parts $4 > 3$, $3< 8$, $8>5$, $5<8=8$, $8 >1$. In both cases we observe the existence of an infinite number of periodic attractors a distance $h\sim p^{\left \lfloor \log_{p} x_{0} \right \rfloor}(p-1)$ apart, each consisting of a continuous band of periodic orbits which is $d\sim p^{\left \lfloor \log_{p} x_{0} \right \rfloor}$ thick. The existence of such periodic attractors is a consequence of the scale invariance of the replacement operator, made explicit by Eq. (\ref{rp0}).

We can now move a step further and consider a slightly more interesting dynamics that is easily understood from the above. Let $m\ge 1$ and $n \ge 2$ be both finite natural numbers. We take
\begin{eqnarray}
j_{t} &\equiv& \left \lfloor \log_{p} x_{t} \right \rfloor-\mathbf{d}_{m}(0,t) \label{mvamos} \\
k_{t} &\equiv& \left \lfloor \log_{p} \mu \right \rfloor-\mathbf{d}_{n}(0,t) \label{nvamos}
\end{eqnarray}
Again, we have two cases depending on whether any of the $n$ most significant digits of $\mu$ in radix $p$ is zero or not. The outcome of the former case is that $x_{t}$ converges to zero for a sufficiently long time, as before. We assume thus the latter case in what follows.
The resulting dynamical evolution depends now on two periods $n$ and $m$ which lead, after a transient, to a periodic orbit, where the period $T=\text{lcm}(m,n)$ is the least common multiple of $m$ and $n$. Note that Eq. (\ref{thedyna}) is trivially invariant under the transformation $t \to t+\text{lcm}(m,n)$ in this case. Again we find attractors with a similar qualitative dynamics albeit with period $T=\text{lcm}(m,n)$, as a consequence of the scale invariance, Eq. (\ref{rp0}).

\subsection{Distinct attractors in phase space with same qualitative dynamics} \label{recipe}

Let us now break the above mentioned scale invariance of Eq. (\ref{rp0}) so that the attractors become distinct. The most straightforward way to achieve this is to make $m$ in Eq. (\ref{mvamos}) dependent on the order of magnitude of $x_{t}$ i.e. on $\left \lfloor \log_{p} x_{t} \right \rfloor$. Since $m \ge 1$ the easiest way to break the invariance is to take 
\begin{eqnarray}
j_{t} &\equiv& \left \lfloor \log_{p} x_{t} \right \rfloor-\mathbf{d}_{m}(0,t) \label{mut1} \\
k_{t} &\equiv& \left \lfloor \log_{p} \mu \right \rfloor-\mathbf{d}_{n}(0,t) \label{nut1} \\
m&\equiv& 1+|\left \lfloor \log_{p} x_{t} \right \rfloor| \label{motl}
\end{eqnarray}
We then note that if $x_{t}$ is finite and its evolution is bounded to an interval $[p^{k}, p^{k+1})$ then $m=k$ is constant, and the orbit has periodicity $\text{lcm}(m,n)=\text{lcm}(k,n)$. Therefore, since $m$ depends on $x_{t}$ we find that the dynamics now has attractors with different periodicities. We note that, since $m=1+|\left \lfloor \log_{p} x_{t} \right \rfloor|=1$ for any initial condition $x_{0} \in [1,p)$, Eq. (\ref{thedyna}) reduces to Eq. (\ref{bmap}). The thickness $d$ calculated previously (see Fig. \ref{oscil1}) is now equal to $d\sim p^{\left \lfloor \log_{p} x_{0} \right \rfloor-m+1}=1$.

In Fig. \ref{oscil2} the evolution of $x_{t}$ as obtained from Eqs. (\ref{thedyna}), (\ref{mut1}), (\ref{nut1}) and (\ref{motl}) for $p=9$ (left panel) and $p=10$ (right panel), for $\mu=2359765$, $n=7$ and $m=1+|\left \lfloor \log_{p} x_{t} \right \rfloor|$ is shown. Besides the oscillations of period $T=7$ for initial conditions $x_{0} \in [1,p)$, which are similar to those in Fig. \ref{oscil1}, oscillations with different periodicities  $T=14, 21, 28$, are now found for sets of initial conditions $x_{0} \in [p, p^{2})$, $x_{0} \in [p^{2}, p^{3})$ and $x_{0} \in [p^{3}, p^{4})$, respectively. Indeed, we have that $T=\text{lcm}\left(m,n\right)$ for the set of initial conditions $x_{0} \in [p^{m}, p^{m+1})$ and these observed periodicities correspond to the fact that $m$ is now dependent on the initial condition through $x_{t}$. Different colors have been chosen in Fig. \ref{oscil2} to indicate the now qualitatively different basins of attraction and the behavior of the initial conditions contained in them, thus emphasizing the breaking of the scale invariance. Although the thickness $d\sim p^{\left \lfloor \log_{p} x_{0} \right \rfloor-m+1}=1$ is independent of the initial condition, the fact that $d$ is finite and does no longer scale with $p^{\left \lfloor \log_{p} x_{0} \right \rfloor}$ is revealed in the logarithmic scale of $x_{t}$ in Fig. \ref{oscil2} at low powers of the radix.

\subsection{Aperiodic nonchaotic attractors}

If $n=\infty$, in Eq. (\ref{nvamos}), we have, by using Eq. (\ref{case2})
\begin{eqnarray}
j_{t} &\equiv& \left \lfloor \log_{p} x_{t} \right \rfloor-\mathbf{d}_{m}(0,t) \label{mSNA} \\
k_{t} &\equiv& \left \lfloor \log_{p} \mu \right \rfloor-t \label{nSNA}
\end{eqnarray}
and Eq. (\ref{thedyna}) reduces to
\begin{equation}
x_{t+1}= \mathcal{R}_{p}\left(\left \lfloor \log_{p} x_{t} \right \rfloor-\mathbf{d}_{m}(0,t),\left \lfloor \log_{p} \mu \right \rfloor-t; x_{t}, \mu \right) \label{aperiodyna}
\end{equation}
We now consider $\mu$ an irrational number (for $\mu$ a rational number, the dynamics can be shown to be equivalent to the one described in Section B, since $\mu$ has then a period of digits after the radix point that repeats infinitely). We have to distinguish two cases: (1) if $\mu$ has $N$ digits equal to zero after the radix point when written in radix $p$, $x_{t}$ can decay up to a factor $p^{-N}$ in the course of its trajectory since, each time such a zero is encountered and used to replace the most significant digit of $x_{t}$ we have $\left \lfloor \log_{p} x_{t+1} \right \rfloor=\left \lfloor \log_{p} x_{t} \right \rfloor-1$, i.e. the position of the most significant digit goes to the next less significant place; (2) if $\mu$ does not contain any digit equal to zero when written in radix $p$, we have $\left \lfloor \log_{p} x_{t} \right \rfloor=\left \lfloor \log_{p} x_{0} \right \rfloor$ i.e., the position of the most significant digit of $x_{t}$ is constant, the dynamics is aperiodic and bounded to the interval $[p^{\left \lfloor \log_{p} x_{0} \right \rfloor},p^{\left \lfloor \log_{p} x_{0} \right \rfloor+1})$ and there exist an infinite number of different aperiodic attractors that are separated and are \emph{not} sensitive to nearby initial conditions. This is easily understood by taking the limit $n \to \infty$ from the case discussed above and illustrated in Fig. \ref{oscil2}. The period of the oscillations is given by $T=\text{lcm}(m,n)$ and when $n$ tends to $\infty$ so does $T$. The dynamics is as follows. The $m$ most significant digits of $x_{t}$ are cyclically replaced by less and less significant digits of $\mu$. Therefore, if $\mu$ is irrational and does not contain any nonzero digit this process does not end on a finite time and, since less significant digits of $\mu$ replace the most significant digits of $x_{t}$ the temporal evolution of $x_{t}$ is necessarily aperiodic. 

To find values of $\mu$ for which no decay to zero happens and one is left with an evolution that is forever aperiodic, let $\alpha$ be any arbitrary irrational number. Then, the number $\mu(\alpha)$ defined by
\begin{equation}
\mu(\alpha)\equiv \frac{p^{1+\left \lfloor \log_{p} \alpha \right \rfloor}}{p-1}+\sum_{j=-\infty}^{\left \lfloor \log_{p} \alpha \right \rfloor}\mathbf{d}_{p-1}(j,\alpha)p^{j} \label{condition}
\end{equation}
is generally irrational as well (for $p>2$) and does not contain any digit equal to zero when written in radix $p$. To see this note that
\begin{eqnarray}
\mu &\equiv& \frac{p^{1+\left \lfloor \log_{p} \alpha \right \rfloor}}{p-1}+\sum_{j=-\infty}^{\left \lfloor \log_{p} \alpha \right \rfloor}\mathbf{d}_{p-1}(j,\alpha)p^{j} \label{condib} \\
&=&\sum_{j=-\infty}^{\left \lfloor \log_{p} \alpha \right \rfloor}(1+\mathbf{d}_{p-1}(j,\alpha))p^{j} 
=\sum_{j=-\infty}^{\left \lfloor \log_{p} \mu \right \rfloor}\mathbf{d}_{p}(j,\mu)p^{j}  \nonumber
\end{eqnarray}
where we have used that $\frac{p^{1+\left \lfloor \log_{p} \alpha \right \rfloor}}{p-1}=\sum_{j=-\infty}^{\left \lfloor \log_{p} \alpha \right \rfloor}p^{j}$. Thus, we find that $\mu$ has digits
\begin{equation}
\mathbf{d}_{p}(j,\mu)=1+\mathbf{d}_{p-1}(j,\alpha) \qquad j \in (-\infty, \left \lfloor \log_{p} \mu \right \rfloor]
\end{equation}
and since $\mathbf{d}_{p-1}(j,\alpha)$ is an integer which obeys $0\le \mathbf{d}_{p-1}(j,\alpha) \le p-2$ then $1\le \mathbf{d}_{p}(j,\mu) \le p-1$, and, therefore, no digit of $\mu$ is equal to zero when written in radix $p$. \emph{That value of $\mu$ leads Eq. (\ref{aperiodyna}) to produce an aperiodic sequence of real numbers which is bounded from above and from below and which is contained in an attractor, the distance between close orbits within the attractor being constant even when the orbit themselves are all aperiodic}. This is so \emph{by mathematical design}. The constancy of the distance shall be rigorously proved below (see Eq. (\ref{lalya})).

\begin{figure*} 
\includegraphics[width=1.0\textwidth]{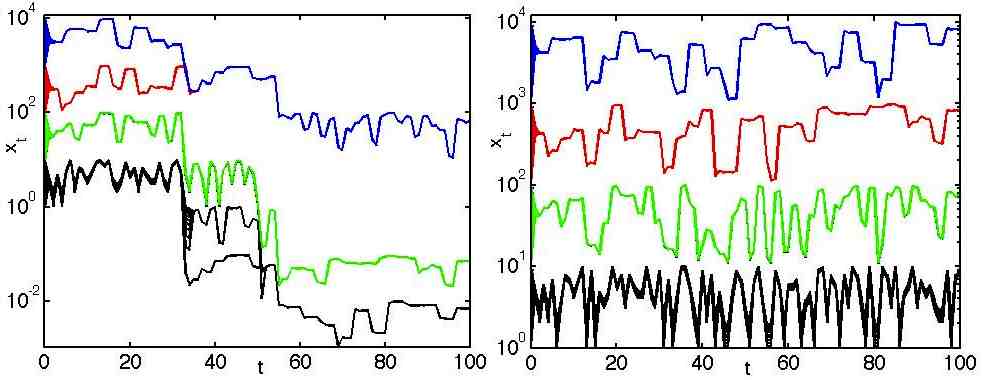}
\caption{\scriptsize{First 100 iterates of Eq. (\ref{aperiodyna2}) for $p=10$ and $\mu=\pi$ (left) and $\mu=\mu(\pi)$ (right), the latter calculated from Eq. (\ref{condib}) for $\alpha=\pi$ and initial conditions $x_{0} \in [1,9900]$.}} \label{aperion}
\end{figure*} 

Before we give a detailed example of this behavior, we mention that the recipe in Section \ref{recipe} to break the scaling invariance also applies in this case if we make $m$ to depend on $t$ so that the dynamics is now given by
\begin{equation}
x_{t+1}= \mathcal{R}_{p}\left(\left \lfloor \log_{p} x_{t} \right \rfloor-\mathbf{d}_{m}(0,t),\left \lfloor \log_{p} \mu \right \rfloor-t; x_{t}, \mu \right) \label{aperiodyna2}
\end{equation}
with
\begin{eqnarray}
j_{t} &\equiv& \left \lfloor \log_{p} x_{t} \right \rfloor-\mathbf{d}_{m}(0,t) \label{mSNA2} \\
k_{t} &\equiv& \left \lfloor \log_{p} \mu \right \rfloor-t \label{nSNA2} \\
m&\equiv& 1+|\left \lfloor \log_{p} x_{t} \right \rfloor| \label{motlS} 
\end{eqnarray}
In this way, all initial conditions contained in each separate interval $[p^{h}, p^{h+1})$ with $h\in \mathbb{Z}$ converge to a distinct separate attractor. 

Let us illustrate all above with a particular example. We focus first on the former case, where $\mu$ may contain zero digits when written in radix $p$. We take $\mu=\pi$
\begin{eqnarray}
\pi&=&3.1415926535897932384626433832795\underline{\mathbf{0}}28841971693 \nonumber \\
&&993751\underline{\mathbf{0}}582\underline{\mathbf{0}}9749445923\underline{\mathbf{0}}78\ldots \nonumber
\end{eqnarray}
and we see that for $p=10$ there are digits equal to zero accompanying powers of the radix $10^{-32}$, $10^{-50}$, $10^{-54}$, $10^{-64}$, etc. Thus, we expect abrupt decays of those  trajectories that satisfy $\mathbf{d}_{1+\left|\left \lfloor \log_{p} x_{t} \right \rfloor\right|}(0,t)=0$, at $t=32, 50, 54, 64,\ldots$. 

In Figure \ref{aperion} (left), the evolution of $x_{t}$ obtained from Eq. (\ref{aperiodyna2}) is plotted for the initial conditions in the set $x_{0}=[1, 9900]$, $m=1+|\left \lfloor \log_{p} x_{0} \right \rfloor|$ and $\mu=\pi$. We see that, although aperiodic trajectories seem to attract subsets of the initial conditions, some of these trajectories abruptly decay at $t=32, 50, 53$ and $64$ which are those times predicted above. For example, the trajectories colored blue (whose initial conditions belong to the interval $[10^{3}, 10^{4})$) satisfy $\mathbf{d}_{1+|\left \lfloor \log_{p} x_{t} \right \rfloor|}(0,t)=\mathbf{d}_{4}(0,32)=0$ at $t=32$ and, therefore, they decay by a factor $p$ and at $t=33$ meet the red curves in the lower level. The latter remain unaffected since they satisfy  $\mathbf{d}_{1+\left|\left \lfloor \log_{p} x_{t} \right \rfloor\right|}(0,t)=\mathbf{d}_{3}(0,32)=2$. At a later time, $t=54$, both blue and red trajectories decay together since now $\mathbf{d}_{1+\left|\left \lfloor \log_{p} x_{t} \right \rfloor\right|}(0,t)=\mathbf{d}_{3}(0,54)=0$. All crossings and complex transitions can thus be explained through the theory presented above. Since $\pi$ may surely contain an infinite number of digits equal to zero (not regularly spaced), we can safely say that \emph{all trajectories decay to zero after a sufficiently long time}.

Let us now construct an irrational number by means of Eq. (\ref{condib}) such that we can explore the case (2) above in which the trajectories do not decay to zero but $x_{t}$ stays bounded within an interval $\left[p^{\left \lfloor \log_{p} x_{0} \right \rfloor}, p^{\left \lfloor \log_{p} x_{0} \right \rfloor+1}\right)$ forever. From Eq. (\ref{condib}) by taking $\alpha=\pi$ we find the irrational number 

\begin{eqnarray}
\mu(\pi)&=&4.25261376469181434957375449438161399521827 \nonumber \\
&& 14114862169311851556134189\ldots \label{mupi}
\end{eqnarray}
which does not contain any zero digit for $p=10$. We now consider the temporal evolution of $x_{t}$ taking this latter number $\mu(\pi)$ as parameter. In Figure \ref{aperion} (right) $x_{t}$ obtained from Eq. (\ref{aperiodyna2}) is plotted for the initial conditions in the set $x_{0}=[1, 9900]$, $m=1+|\left \lfloor \log_{p} x_{0} \right \rfloor|$, $p=10$ and $\mu=\mu(\pi)$. Stable aperiodic oscillations are shown to attract connected sets of initial conditions. The attractors are all qualitatively distinct, by virtue of the scaling symmetry breaking, and each displays a distinct aperiodic behavior. The trajectory exhibits more abrupt and sudden changes for $m=1+|\left \lfloor \log_{p} x_{0} \right \rfloor|$ close to unity (i.e. for $x_{0} \in [1,p)$) and is coarser for $x_{0}$ larger. We observe that there is no exponential sensitivity to initial conditions even when the resulting dynamics is forever aperiodic: all close initial conditions within an interval $\left[p^{\left \lfloor \log_{p} x_{0} \right \rfloor}, p^{\left \lfloor \log_{p} x_{0} \right \rfloor+1}\right)$ are attracted to a thin band of aperiodic trajectories of thickness $d \sim 1$ where they keep a constant distance.

We can, indeed, rigorously prove that the distance of trajectories starting infinitesimally close remains constant: The Lyapunov exponent of the map is zero for every possible value of $m$ in Eq. (\ref{motlS}) as we proceed to show. Let $x_{t}$ and $x_{t}'$ be two different trajectories starting from initial conditions $x_{0}$ and $x_{0}'$ that are infinitesimally close. In practice, this means that we can take 
\begin{equation}
\Delta_{0}\equiv |x_{0}-x_{0}'|=\sum_{k=-\infty}^{-s-1} p^{k} \mathbf{d}_{p}(k,|x_{0}-x_{0}'|) < p^{-s}\label{zerote}   
\end{equation}
with $s$ sufficiently small. Here we have used Eq. (\ref{iden}) to expand $|x_{0}-x_{0}'|$ in radix $p$ thus making clear that the digits of the initial conditions $x_{0}$ and $x_{0}'$ are equal up to $s$ significant digits after the radix point. Since $m$ is equal for both trajectories that are infinitesimally separated because $m=1+|\left \lfloor \log_{p} x_{0} \right \rfloor|=1+|\left \lfloor \log_{p} x_{0}' \right \rfloor|$ and since $m$ is finite, positive, and non-vanishing, it is enough to take $s > m$. The dynamics only affects the $m$ most significant digits of $x_{t}'$ and $x_{t}$ \emph{and leaves all other unaffected}. Furthermore, and crucially, each single digit of $x_{t}$ and $x_{t}'$ that is being replaced at each time $t$ is \emph{at the same position} relative to the radix point for both numbers, and it is replaced \emph{by the same digit} of $\mu$ (i.e. the $k_{t}$-th digit of $\mu$, as given by Eq. (\ref{nSNA2}) at each $t$). Consequently, the difference of the contributions coming from these $m$ digits to $|x_{t}-x_{t}'|$ cancel out and we are only left with
\begin{eqnarray}
\Delta_{t}&\equiv& |x_{t}-x_{t}'|=\sum_{k=-\infty}^{-s-1} p^{k} \mathbf{d}_{p}(k,|x_{t}-x_{t}'|) \nonumber \\
&=& \sum_{k=-\infty}^{-s-1} p^{k} \mathbf{d}_{p}(k,|x_{0}-x_{0}'|)=\Delta_{0} \label{zerotet}   
\end{eqnarray}
 at every time. Therefore, the Lyapunov exponent 
\begin{eqnarray}
\lambda &=& \lim_{t\to \infty} \frac{1}{t}\ln \frac{\Delta_{t}}{\Delta_{0}}=\lim_{t\to \infty} \frac{1}{t}\ln \frac{\sum_{k=-\infty}^{-s-1} p^{k} \mathbf{d}_{p}(k,|x_{0}-x_{0}'|)}{\sum_{k=-\infty}^{-s-1} p^{k} \mathbf{d}_{p}(k,|x_{0}-x_{0}'|)} \nonumber \\
&=& 0 \label{lalya}
\end{eqnarray}
which means that the attractor, \emph{in spite of being aperiodic, is not chaotic}. The above proof is equally valid for the dynamics described in Sections B and C. For the dynamics in Section A we, trivially, find $\lambda=-\infty$, since $\Delta_{t} \to 0$ as $t \to \infty$ as a consequence of the existence of a stable fixed point given by Eq. (\ref{fpoints}) which is the same for both trajectories.

We must emphasize the importance of $\mu$ being \emph{irrational} for the dynamics above being aperiodic. Because if $\mu$ were rational, then it would have an infinitely repeating period of, say, $a_{0}$ digits after the radix point, and the dynamics would trivially reduce to the ones described in Sections B and C (with $n=a_{0}$ finite).

Let us summarize the results of the above analysis, all trivial consequences of our construction:
\begin{itemize}
\item The dynamics is aperiodic on the attractor.
\item The Lyapunov exponent is zero.
\item This behavior is possible only if $\mu$ is an irrational number. If $\mu$ is rational, the dynamics is periodic. 
\end{itemize}

Although we are not establishing here the fractal character of the aperiodic attractors, we must remark the analogy of the behavior that we have described with the one found in so-called \emph{strange nonchaotic attractors} (SNAs)\cite{Grebogi, Feudel1, Feudel2, HuntO, Yalcinkaya, Prasad,Ramas1,Ramas2,Ramas3}, which are driven by a forcing consisting of an incommesurate ratio of frequencies (this ratio being an irrational number that plays an analogous role to $\mu$ here).  A SNA is a fractal structure attracting the trajectories in phase space so that the dynamics of the system is not exponentially sensitive to initial conditions (its Lyapunov exponent being non-positive). The existence of SNAs was first pointed out in a seminal paper by Grebogi, Ott, Pelikan and Yorke \cite{Grebogi} and have been widely studied since \cite{Feudel1, Feudel2, HuntO, Yalcinkaya, Prasad,Ramas1,Ramas2,Ramas3}. Experimentally, SNAs were first discovered in the buckling of a magnetoelastic ribbon driven quasiperiodically by two incommensurate frequencies with the golden ratio relationship \cite{Ditto}. Since then they have also been reported in other laboratory experiments involving electrochemical cells \cite{Parmananda}, electronic circuits \cite{Zhou}, and a neon glow discharge \cite{Ding}. Recently, based on the unprecedented light curves of the Kepler space telescope, SNAs have also been observed for the first time out of the laboratory in the dynamics of the RRc Lyrae star KIC 5520878, a blue-white star 16 000 light years from Earth in the constellation Lyra \cite{Lindner}. 

The temporal dynamics exhibited by the trajectories in Fig. \ref{aperion} (right) on any of the attractors (differently colored) seems very similar to the time series found in SNAs, as for example the ones obtained (in continuous time) with a modified driven pendulum equation that has been used to model a driven SQUID with inertia and damping\cite{Zhou}, see also Fig 1c in\cite{Ramas2}. We must however remark that the dynamics is here time-discrete and that the Lyapunov exponent is zero. 

If we set $m$ constant, independent of $x_{t}$ (i.e. $x_{0}$) in Eq. (\ref{motlS}), we recover back the scaling invariance and the above model reduces to Eqs. (\ref{mSNA}), (\ref{nSNA}) and (\ref{aperiodyna}) and all aperiodic attractors in phase space become synchronized in phase (which in this context means that, although the signals $x_{t}$ have different value, all their local maxima and minima are locked and synchronous in time). Thus, in this situation, if one thinks on each separate aperiodic attractor as reflecting the evolution of an independent subsystem within a composite system, the aperiodic attractors are synchronized very much like in reference \cite{Ramas1}, where synchronization of SNAs is discussed.

Although we must conclude here our discussion of Eq. (\ref{thedyna}) it should now be clear that if $\mu$ is no longer a parameter but a function of $x_{t}$ as well, we can have chaotic attractors as a result. Already if we simply replace $\alpha$ by $x_{t}$ in Eq. (\ref{condib}) and $\mu$ by $\mu(x_{t})$ in Eq. (\ref{aperiodyna}) or (\ref{aperiodyna2}), the result is an aperiodic dynamics that is exponentially sensitive to initial conditions: The closer two different initial conditions $x_{0}$ and $x_{0}'$ are, the longer the trajectories evolve together. However, as soon as the corresponding digits of $x_{t}$ and $x_{t}'$ that are being used to replace the $m$ most significant digits of $x_{t}$ and $x_{t}'$ begin to differ, the trajectories begin to depart aperiodically from each other and one can show that the Lyapunov exponent is positive in this case. In the long term, the value of $x_{t}'$ cannot be estimated from the value of $x_{t}$ even when both $x_{t}$ and $x_{t}'$ were initially close.

\section{Conclusions}

In this work we have applied methods of digital mathematics \cite{QUANTUM, VGM4, CHAOSOLFRAC, PHYSAFRAC} to construct a generic nonlinear map which operates by replacing digits of the numerical value of a dynamical variable $x_{t}$ by those of a control parameter (or signal) $\mu$. By considering increasingly complex dynamics, we have shown how our map can be used to mathematically design aperiodic non-chaotic attractors. We have explicitly shown how this map can be understood as a discretization of any arbitrary non-autonomous map by working out in detail a mathematical method to achieve this connection in general. Then, we have investigated the most simple cases of digit replacement dynamics. Although the notation and the functions introduced here may seem unfamiliar, we believe that the nonlinear behavior obtained from the map is simpler to understand than the complex behavior arising from polynomial normal forms and maps involving transcendental functions. We have explicitly shown how limit cycle oscillations and aperiodic behavior can be mathematically designed, showing the advantages of this formulation. 

The theoretical results presented in the manuscript may be reproducible in experiments with nonlinear electric circuits. The digit function can be obtained in practice from a certain superposition of square waves or from a sawtooth signal appropriately quantized. Thus, a circuit element performing the replacement operator may be constructed and the electric current through it may be shown to reproduce the results presented in the manuscript.

Although all physical numbers are radix independent, they are all always \emph{represented} in a specific radix (e.g. the usual decimal radix). We have here thus constructed a new class of dynamical systems by regarding the possibility that the digits of the numbers are being replaced at each instant of time (what every function indeed does) one by one. Note that in this article it is \emph{irrelevant} which radix we actually choose for giving the numbers. However, if we choose the optimal \emph{representation} for the radix $p$ at which the digits are being replaced, it becomes at once clear how dynamical processes, obtained through digit replacement, evolve in time.


\begin{thebibliography}{10}

\bibitem{May}
R.~May,
\newblock Nature {\bf 261}, 459 (1976).

\bibitem{PeitgenBOOK}
H.~O. Peitgen, H.~J\"urgens, and D.~Saupe,
\newblock {\em Chaos and Fractals} (Springer Verlag, New York, 2004).

\bibitem{Lichtenberg}
A.~J. Lichtenberg and M.~A. Lieberman,
\newblock {\em Regular and chaotic dynamics} (Springer, New York, 2010).

\bibitem{Wolfram}
S.~Wolfram,
\newblock {\em A New Kind of Science} (Wolfram Media Inc., Champaign, IL,
  2002).

\bibitem{Lorenz}
E.~N. Lorenz,
\newblock J. Atmos. Sci. {\bf 20}, 130 (1963).

\bibitem{Rossler}
O.~E. R\"ossler,
\newblock Phys. Lett. A {\bf 57}, 397 (1976).

\bibitem{Feigenbaum}
M.~J. Feigenbaum,
\newblock J. Stat. Phys. {\bf 19}, 25 (1978).

\bibitem{CHAOSOLFRAC}
V.~Garc{\'\i}a-Morales,
\newblock Chaos Sol. Fract. {\bf 83}, 27 (2016), cond-mat/1505.02547v3.

\bibitem{PHYSAFRAC}
V.~Garc{\'\i}a-Morales,
\newblock Physica A {\bf 447}, 535 (2016), cs.OH/1507.01444v3.

\bibitem{semipredo}
V.~Garc{\'\i}a-Morales,
\newblock Commun. Nonlinear Sci. Numer. Simulat. {\bf 39}, 81 (2016).

\bibitem{EPL}
V.~Garc{\'\i}a-Morales,
\newblock EPL {\bf 114}, 18002 (2016).

\bibitem{Lind}
D.~Lind and B.~Marcus,
\newblock {\em Symbolic Dynamics and Coding} (Cambridge University Press,
  Cambridge, UK, 1995).

\bibitem{RuelleEckmann}
J.-P. Eckmann and D.~Ruelle,
\newblock Rev. Mod. Phys. {\bf 57}, 617 (1985).

\bibitem{VGM4}
V.~Garc{\'\i}a-Morales,
\newblock Physica A {\bf 440}, 110 (2015).

\bibitem{QUANTUM}
V.~Garc{\'\i}a-Morales,
\newblock Found. Phys. {\bf 45}, 295 (2015).

\bibitem{Indekeu5}
J.~O. Indekeu, G.~Fleerackers, A.~I. Posazhennikova, and E.~Bervoets,
\newblock Physica A {\bf 285}, 135 (2000).

\bibitem{Indekeu4}
A.~I. Posazhennikova and J.~O. Indekeu,
\newblock Int. J. Thermophys. {\bf 22}, 1123 (2001).

\bibitem{Indekeu6}
A.~I. Posazhennikova and J.~O. Indekeu,
\newblock Physica A {\bf 414}, 240 (2014).

\bibitem{StanleyFractals}
A.~L. Barabasi and H.~E. Stanley,
\newblock {\em Fractal concepts in surface growth} (Cambridge University Press,
  Cambridge, UK, 1995).

\bibitem{Barnsley}
M.~F. Barnsley,
\newblock {\em Superfractals} (Cambridge University Press, Cambridge, UK,
  2006).

\bibitem{Grebogi}
C.~Grebogi, E.~Ott, S.~Pelikan, and J.~A. Yorke,
\newblock Physica D {\bf 13}, 261 (1984).

\bibitem{Feudel1}
A.~Pikovsky and U.~Feudel,
\newblock Chaos {\bf 5}, 253 (1995).

\bibitem{Feudel2}
U.~Feudel, S.~Kuznetsov, and A.~Pikovsky,
\newblock {\em Strange Nonchaotic Attractors: Dynamics between Order and Chaos
  in Quasiperiodically Forced Systems} (World Scientific, Singapore, 2006).

\bibitem{HuntO}
B.~R. Hunt and E.~Ott,
\newblock Phys. Rev. Lett. {\bf 87}, 254101 (2001).

\bibitem{Yalcinkaya}
T.~Yalcinkaya and Y.~C. Lai,
\newblock Phys. Rev. Lett. {\bf 77}, 5039 (1996).

\bibitem{Prasad}
A.~Prasad, V.~Mehra, and R.~Ramaswamy,
\newblock Phys. Rev. Lett. {\bf 79}, 4127 (1997).

\bibitem{Ramas1}
R.~Ramaswamy,
\newblock Phys. Rev. E {\bf 56}, 7294 (1997).

\bibitem{Ramas2}
A.~Prasad, S.~S. Negh, and R.~Ramaswamy,
\newblock Int. J. Bifurcation Chaos {\bf 11}, 291 (2001).

\bibitem{Ramas3}
A.~Prasad, A.~Nandi, and R.~Ramaswamy,
\newblock Int. J. Bifurcation Chaos {\bf 17}, 3397 (2007).

\bibitem{Ditto}
W.~Ditto {\em et~al.},
\newblock Phys. Rev. Lett. {\bf 65}, 533 (1990).

\bibitem{Parmananda}
G.~Ruiz and P.~Parmananda,
\newblock Phys. Lett. A {\bf 367}, 478 (2007).

\bibitem{Zhou}
T.~Zhou, F.~Moss, and A.~Bulsara,
\newblock Phys. Rev. A {\bf 45}, 5394 (1992).

\bibitem{Ding}
W.~Ding, H.~Deutsch, A.~Dinklage, and C.~Wilke,
\newblock Phys. Rev. E {\bf 55}, 3769 (1997).

\bibitem{Lindner}
J.~F. Lindner {\em et~al.},
\newblock Phys. Rev. Lett. {\bf 114}, 054101 (2015).

\end{thebibliography}
\end{document}